# Strain induced variations in transport and optical properties of SrVO$_3$: a DFT+U study

Maitreyo Biswas[1], Debolina Misra[2,*] and Tarun K. Kundu[1]

[1]Department of Metallurgical and Materials Engineering, Indian Institute of Technology Kharagpur, India, 721302

[2]Department of Physics, Indian Institute of Information Technology, Design and Manufacturing, Kancheepuram, Chennai, India, 600127.

*debolinam@iiitdm.ac.in

## Abstract:

First-principles calculations based on density functional theory + Hubbard U (DFT+U) approach have been carried out to study the strain induced variations in the optical and transport properties of the correlated perovskite SrVO$_3$. By virtue of its conductivity, high carrier mobility and optical transparency, SrVO$_3$ can be used as a potential replacement of indium tin oxide (ITO) as a transparent conductor. As strain tuning is an effective way to tune the electron-electron correlations in correlated oxides, the epitaxial strain induced variations in V-3$d$ bandwidth, band center shift and band splitting at high symmetry points (Γ, R) in SrVO$_3$ are investigated. The alterations in resistivity, carrier concentration, Hall coefficient and plasma frequency with applied strain are also elucidated. Our calculations revealed that under tensile strain, the lifting of the threefold degeneracy of 3$d$-$t_{2g}$ orbital and $d$-band narrowing reinforces a relatively less conducting state thus limiting the ω$_P$ to lower frequencies. On the contrary, in case of compressive strain the $d$-band widening predominates leading to an increase in carrier concentration and decrease in resistivity enhancing the metallic state. As a result, ω$_P$ is increased to higher frequencies which decreases the optical transparency window. Hence, our results and findings clearly demonstrate the interdependence between the optical and transport properties, and provides a detailed mechanism to tune the optoelectronic properties of SrVO$_3$ for its applications as a transparent conducting oxide.

## 1. Introduction:

Correlated metal oxides have been a field of extensive study as they exhibit remarkable electronic, optical, magnetic, ferroelectric and piezoelectric properties [1-5]. These oxides show intriguing properties owing to the narrow transition-metal (TM) $d$-band and high electron-electron correlations. Among these correlated oxides, SrVO$_3$ (SVO) has drawn major attention recently for a wide range of potential applications especially as a transparent conducting oxide (TCO) [6]. The room temperature carrier concentration of SVO is reported to be ~$2 \times 10^{22}$ cm$^{-3}$ with a plasma frequency $\hbar\omega_p = 1.33\ eV$ [6-8]. For applications of SVO in different fields and with desired efficiency, tuning its properties by strain engineering has emerged as a very important tool as it provides an effective way to precisely tune the band gaps, band structures, optical properties and transport properties of perovskites [9-12].

The examples of strain tuning of SVO include its growth on SrTiO$_3$ [7,13], (LaAlO$_3$)$_{0.3}$(Sr$_2$AlTa$_6$)$_{0.7}$ (LSAT) [7,14], LaAlO$_3$ [15] using PLD and hybrid-MBE and subsequent lowering of the point group symmetry to $D_{4h}$ similar to SrTiO$_3$ [10,16] and LaVO$_3$ [17]. The strain induced variations in the transport properties and plasma frequency ($\omega_P$) of SVO films as well as bulk SVO have been analyzed in various experimental and theoretical studies [2,7,18-20]. A thorough understanding of the interplays between the strain induced band splitting, V-3$d$ bandwidth alterations, transport and optical property tuning for the potential application of SVO as a TCO have not been explored intensively so far.

Hence, employing the generalized gradient approximation + U (GGA+U) approach, we intend to provide a clear insight into the variations in the optical and transport properties of SVO under the application of strain. Density of states, V-3$d$ bandwidth and band-center shift, carrier concentration, electrical resistivity and plasma frequency are calculated to clearly reveal how epitaxial strain effectively controls the electronic correlations for device applications.

## 2. Computational details:

First-principles calculations based on density functional theory (DFT) as implemented in the Vienna Ab-initio Simulation Package (VASP) [21] in the MedeA-VASP software, were carried out for both unstrained and strained SrVO$_3$. Unstrained SrVO$_3$ has a centrosymmetric cubic structure ($Pm\bar{3}m$) [17,22] with a five-atom primitive cell, where the central V atom is surrounded by six O-atoms forming VO$_6$ octahedra and having $O_h$ point group symmetry [18]. As SrVO$_3$ is known to be a strongly correlated oxide [17], we moved beyond the conventional DFT approaches and adapted the DFT+U method [23] for an accurate description of its ground state. The GGA-PBE functional [24] with an effective U ($U_{eff}$) value of 2.3 eV [25] has been used for all our calculations; here $U_{eff}$ = U – J; U and J refer to the onsite Hubbard energy and the Hund's coupling terms respectively [26,27].

Structural optimizations were carried out employing the projector-augmented wave (PAW) method [26] and the conjugate gradient algorithm [28] with a 6×6×6 Γ-centered k-mesh, and a plane wave cutoff of 520 eV. The atoms were allowed to relax fully during the optimization process with an energy convergence of 10$^{-5}$ eV and a force convergence of 0.02 eV/Å$^2$ for each atom. The density of states (DOS) calculations were carried using denser k-grids and employing the tetrahedron method [23] for Brillouin-zone integration.

The Hall coefficient and resistivity had been calculated using the Boltztrap code [29,30] which is based on the semi-classical Boltzmann theory of carrier transport. In the Boltztrap code, the transport tensor is expressed in terms of relaxation time and group velocity, and ultimately obtained as a function of chemical potential and temperature [31]. In addition, the Fourier expansion of the band-energies is performed and the electrical conductivity is calculated as σ/$\tau$, where σ denotes the conductivity tensor and $\tau$ constant relaxation time. To investigate the trends in electrical conductivity, the constant relaxation time approximation [30,32] had been used. From the analytical representation of bands, as used in Boltztrap, the Hall coefficient is calculated from the second derivative of the bands. Under the approximation that relaxation time $\tau$ is band index and direction independent, Hall coefficient is found to be independent of $\tau$.

## 3. Results and Discussion:

Vanadium atom (V) in SrVO$_3$, is in 4+ oxidation state with a partially filled $t_{2g}$ orbital and an empty $e_g$ orbital ($t_{2g}^1 e_g^0$) [17,25]. The lattice parameter a$_0$ = 3.89 Å, and the Sr—O (2.74 Å) and V—O (1.94 Å) bond-lengths obtained for the unstrained SrVO$_3$ from our DFT calculation match excellently with previous reports [25,33].

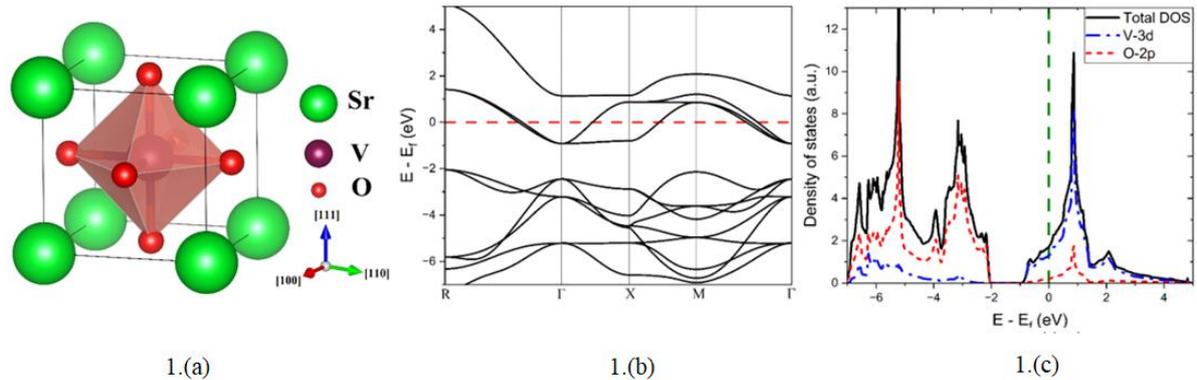

*Figure 1: The (a) optimized unit cell (b) band dispersion and (c) density of states of unstrained SrVO$_3$*

From the bandstructure calculations, shown in Fig:1 (b), it is evident that the in cubic SVO, both the valence band maxima (VBM) and the conduction band minima (CBM) possess threefold degeneracy. While the CBM is situated at the Γ point, VBM lies at the high symmetry point R and the Fermi level (dotted line) passes through the conduction band (Fig.1(c)). The CBM consisting almost entirely of V-3d states is 1.17 eV away from the VB, which is consisted of the O-2*p* states [17].

Effect of biaxial strains have been simulated through alternation of the in-plane lattice parameters a$_0$ and b$_0$ as per the relation $a_f = a_0 + \varepsilon a_0$, where a$_0$ is the unstrained lattice parameter and a$_f$ is the strained lattice parameter. Here ε is the strain parameter which varies from -5% to +5%, where, ε < 0 represents compressive and ε > 0 denotes tensile biaxial strain. For each of these structures, the corresponding optimal out-of-plane lattice parameter (*c*) was calculated using the relation, $c = \left[1 - \frac{2\nu}{1-\nu}\varepsilon\right]a_0$ [14]; $\nu$ being the Poisson ratio. For our calculations, the Poisson ratio, $\nu$ = 0.24 was used as mentioned in a recent experimental study [22].

### 3.1 Changes in electronic structure due to strain:

*Figure 1.b,c*, reflects the metallic nature of unstrained SrVO$_3$ as the Fermi level passes through the conduction band having non-vanishing density of states. As shown in *Figure 2*, the metallic nature is retained over a wide range of biaxial compressive and tensile strains. The variations in the conduction bandwidth and band separation observed in *Figure 2* has been explained in detail in the following sections.

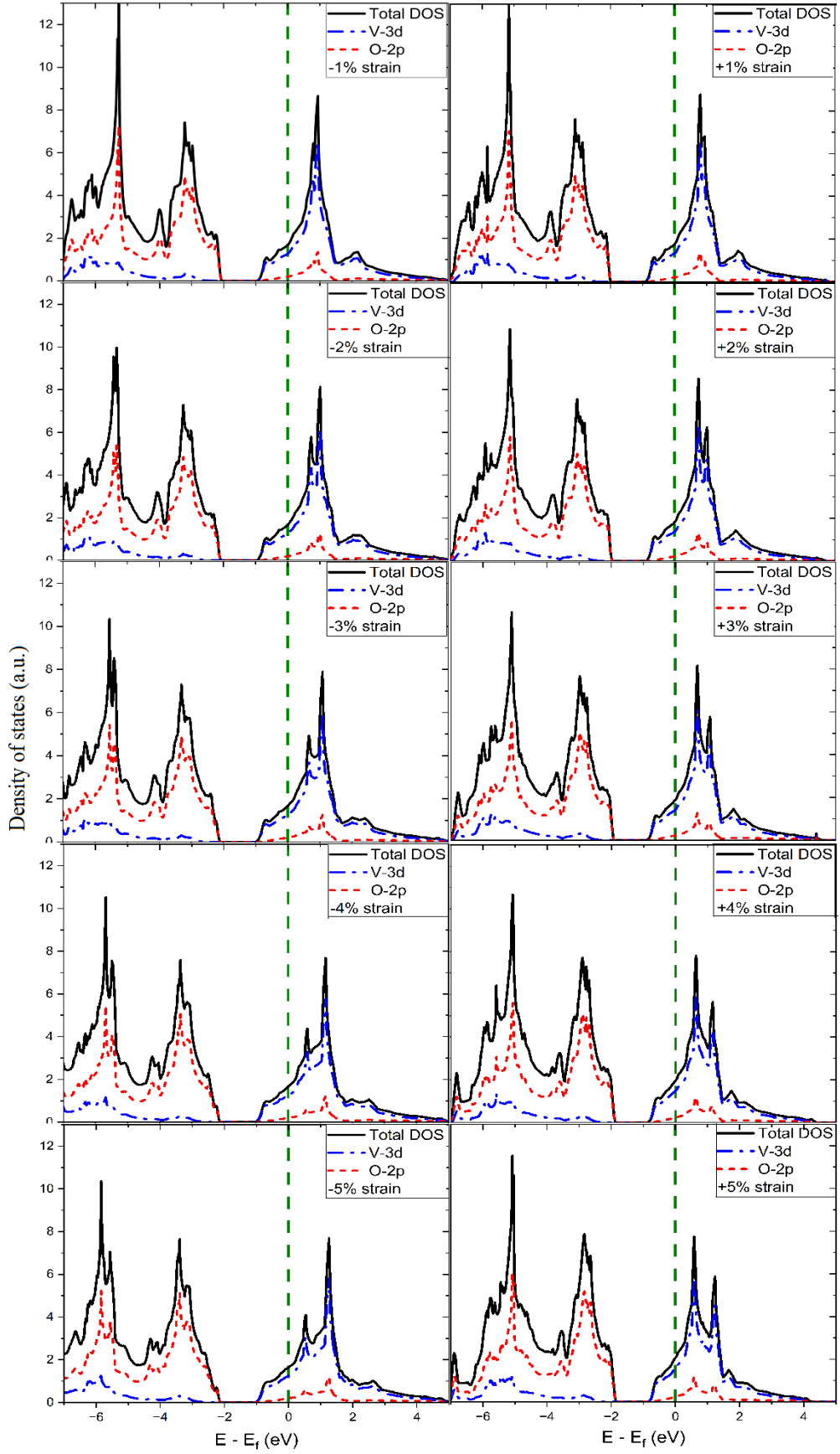

*Figure 2: Total and projected DOS of SVO under tensile and compressive strain*

## 3.2 Strain induced variations in V-*d* bandwidth ($W_d$) and shifts in *d*-band center:

As SrVO$_3$ is a strongly correlated oxide, to understand the effect of strain on its electronic properties, the V-*3d* band in detail has been studied. The d-band center ($e_d$) was calculated as the first moment of the projected density of states of the V-d states with respect to the fermi level ($E_f$) whereas the bandwidth ($W_d$) was calculated as the square root of the second moment of the projected density of states of the V-*3d* states with respect to the fermi level ($E_f$) [34,35].

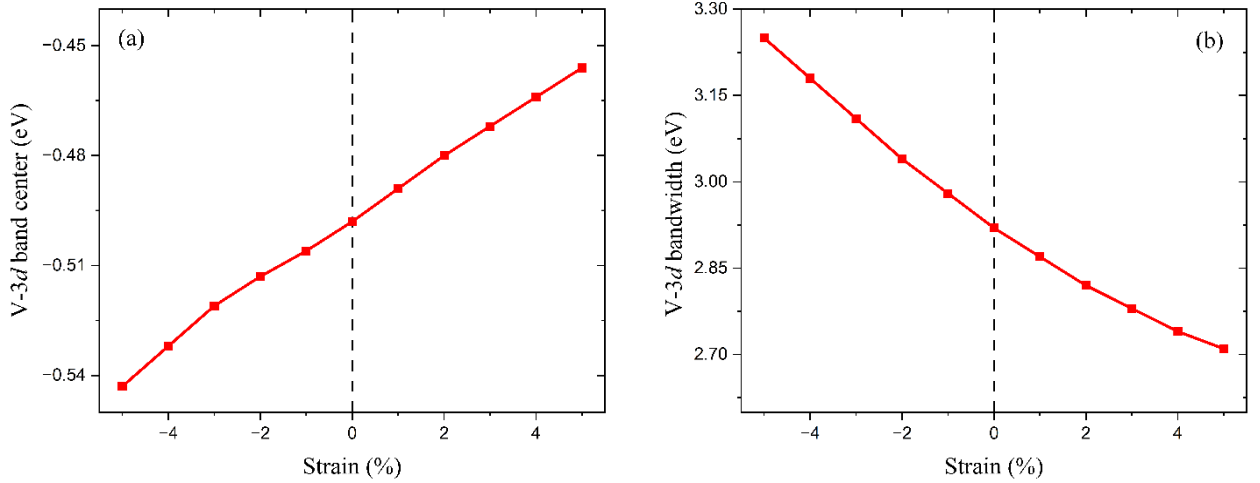

*Figure 3: Variation of the (a) V-3d band-center and (b) V-3d bandwidth with applied biaxial strain*

Strain has been observed not to affect the electronic properties of SVO strongly. However, from *Figure 3.a*, one can observe that while the tensile strain shifts the *d*-band center ($e_d$) towards the Fermi level whereas $e_d$ is shifted away from the Fermi level under compressive strain. This trend can be explained using the rectangular *d*-band model [35,36]: when the *d*-band widens, to maintain a constant *d*-band filling the *d*-band center moves down in energy away from the Fermi level. An exactly opposite trend can be observed when the *d*-band narrows down and consequently the *d*-band center moves up in energy, towards the Fermi level. It can hence be expected that the *d*-band widens under compressive strain and narrowing of *d*-band takes place under tensile strain. To verify this, we have calculated and analyzed the V-*3d* bandwidth ($W_d$) variations under strain. From *Figure 3.b* it is evident that the V-*3d* bandwidth increases under compressive strain and narrows down under tensile strain. In such bulk systems, this can be attributed to the relative changes in orbital overlap due to changing nearest neighbour distances between O and the central V-atom [36]. Previously, similar trends have been observed in metallic systems where the *d*-band center and *d*-bandwidth were linearly correlated [35,37]. Thus, the variation in bandwidth ($W_d$) although not high, agrees with the rectangular band model.

In addition, due to tensile strain, the ratio $U/W_d$ ($U$ – on-site Columbic repulsions) increases with decreasing $W_d$, thus causing higher electron-electron correlations and electron localizations indicating the relatively less conducting phase. On the other hand, *d*-band widening due to compressive strain decreases the $U/W_d$ ratio, which leads to a decrease in localizations and effective electron mass, thereby enhancing the metallic behaviour as compared to unstrained SVO due to decreasing electron-electron correlations.

## 3.3 Hall coefficient and carrier concentration:

The Hall coefficient and carrier concentration at 300 K, at different strain values have been listed in *Table 1*. Carrier concentration (*n*) can be estimated from the Hall coefficient ($R_H$) using the equation: $R_H = \frac{-1}{ne}$. The negative value of $R_H$ implies that the majority carriers are electrons.

*Table 1: Variation of Hall coefficient ($R_H$) and carrier concentration (n) with applied biaxial strain*

| Biaxial strain (%) | Hall coefficient ($R_H \times 10^{-10}$, m$^3$/C) | Carrier concentration ($n \times 10^{22}$, cm$^{-3}$) | V-O bond length (in-plane direction, Å) | V-O bond length (out-of-plane direction, Å) |
|---|---|---|---|---|
| -5 | -1.47 | 4.25 | 1.85 | 2.00 |
| -4 | -2.43 | 2.58 | 1.87 | 1.99 |
| -3 | -2.50 | 2.50 | 1.88 | 1.98 |
| -2 | -2.54 | 2.46 | 1.90 | 1.97 |
| -1 | -2.60 | 2.41 | 1.92 | 1.96 |
| 0 | -2.64 | 2.37 | 1.94 | 1.94 |
| 1 | -2.68 | 2.34 | 1.96 | 1.93 |
| 2 | -2.75 | 2.28 | 1.98 | 1.92 |
| 3 | -2.77 | 2.26 | 2.00 | 1.91 |
| 4 | -2.78 | 2.25 | 2.02 | 1.89 |
| 5 | -2.79 | 2.24 | 2.04 | 1.88 |

At room temperature, bulk SrVO$_3$ showed the carrier concentration of ~$2.34 \times 10^{22}$ cm$^{-3}$, which is comparable to the reported experimental values [6,7,38]. The carrier concentration has been observed to increase with the increasing value of compressive strain and decreases gradually as tensile strain is applied. The decrease in carrier concentration under tensile strain is attributed to the relative localization (with respect to the unstrained state) caused by the increased electron-electron correlation because of decreasing V-*3d* bandwidth. An opposite trend is observed in case of compressive strain where the carrier concentration increases due to increasing V-*3d* bandwidth.

## 3.4 Resistivity:

For calculating resistivity, a fixed constant relaxation time approximation ($\tau$) in the range of ~fs (10$^{-15}$ s) was used. Similar assumptions have been used earlier for studying the thermoelectric properties of SrVO$_3$/SrTiO$_3$ superlattices at 300 K [39,40] as well as for SrTiO$_3$ [40] that has the same point group symmetry (*Pm$\bar{3}$m*) as of unstrained SVO.

At room temperature (300 K), the resistivity of SrVO$_3$ ($\rho_{300}$) is found to be 30.41 µΩ cm which is in well agreement with the experimental value reported (30-40 µΩ cm) [7,42]. *Figure 4* shows the variation of resistivity at room temperature under biaxial strain.

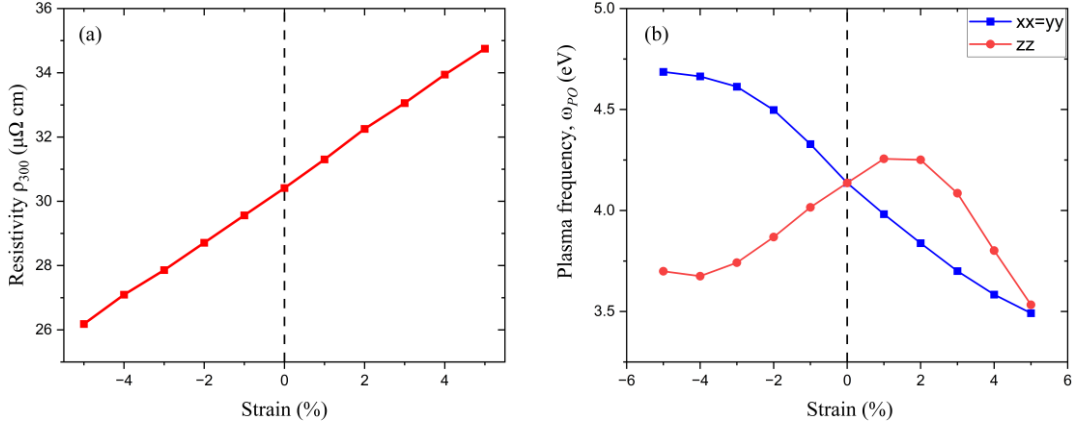

*Figure 4: Variation of (a) Resistivity ($\rho_{300}$) at 300 K and (b) Plasma frequency ($\omega_{PO}$) with the applied strain*

Similar to electronic structure, the resistivity value is also not affected substantially under strain. However, it has been observed that on applying tensile strain, the resistivity increases, whereas opposite trend is observed under compressive strain. Under tensile strain, the carrier concentration decreases (*Section 3.3*) and electron-electron correlations increases due to decrease in V-*3d* bandwidth (*Section 3.2*). The increase in electron localizations due to increased correlations play a major role in increasing the resistivity. Application of compressive strain shows the opposite trends and thus resistivity decreases with increasing compressive strain.

### 3.5 Effect of strain on plasma frequency:

Plasma frequency of a material is very important to estimate its transparency window [2,20]. Under the DFT+U framework we have calculated the screened plasma frequency ($\omega_p$) which is defined as: $\hbar\omega_P = \sqrt{\frac{ne^2}{\varepsilon_r m^*}}$, where *n, m\** and $\varepsilon_r$ are the carrier concentration, effective mass, and relative permittivity of electrons respectively [20,43]. The unscreened plasma frequency ($\omega_{PO}$) is defined as the square root of the Drude weight and hence depends only on $\sqrt{\frac{n}{m^*}}$ [43]. As $\omega_{PO}$ is independent of the screening by the core electrons, we have calculated $\omega_{PO}$ to understand how the optical transparency of the material changes with carrier concentration (*n*) and electron localizations. Finally, $\omega_{PO}$ and $\omega_P$ are related via the equation: $\omega_{PO} = \omega_P\sqrt{\varepsilon_{core}}$ where $\varepsilon_{core}$ is the relative permittivity of core electrons [20].

*Figure 4.b* helps to understand how the variation of electron-electron correlations due to strain affects the plasma frequency and the transparency window.

The unscreened plasma frequency of unstrained SVO obtained from our calculation is isotropic in nature with $\omega_{PO}$ = 4.14 eV which supports the previously reported value of the plasma frequency of ~4.18 eV [20]. As the biaxial strain distorts the cubic symmetry and changes it from *Pm$\bar{3}$m to P4/mmm*, the plasma frequency becomes anisotropic with (($\omega_{PO}$)xx = ($\omega_{PO}$)yy ≠ ($\omega_{PO}$)zz. The plasma frequency shows opposite trends along the two directions: in-plane and out-of-plane as both directions experience opposite nature of strain. However, there is a dip in the value of out-of-plane plasma frequency after 2% tensile strain.

An enhanced effective mass of electron ($m^*$) under tensile strain owing to increasing electron-electron correlations, and a decrease in the carrier concentration ($n$) results in a decrease in the plasma frequency $\omega_{PO}$ as observed in Figure 4b. For the compressive strain, on the contrary, the opposite situation prevails and $\omega_{PO}$ increases with increasing compressive strain.

The increase in the in-plane plasma frequency value at higher values of compressive strain indicates that strained SVO possesses high reflectivity upto higher frequencies of light which limits the transparency window. Under tensile strain, on the other hand, the in-plane plasma frequency value decreases, and hence, an increase in the transparency window is expected. Our results hence clearly show that a trade-off exists between conductivity, carrier concentration and transparency window, and a careful tuning is necessary in order to achieve the desired efficiency of the material.

The decrease in out-of-plane plasma frequency value above 2% tensile strain is attributed to the reinforcement of the electron localizations and electron-electron correlations due to $t_{2g}$-band splitting and *d*-band narrowing (as discussed in detail in *Section 3.6*). Under the influence of high tensile strain, the electron-electron correlations can increase to such an extent that even in the out of plane direction the electrons get localized which forces the $(\omega_{PO})_{zz}$ to attain an isotropic nature, as observed for 5% tensile strain (Figure 4b).

### **3.6 Band splitting at high symmetry points:**

In unstrained cubic SVO we observed threefold degeneracy at the CBM(Γ) and VBM(R) (Figure 1.b) and a point group symmetry of $O_h$. The application of strain changes the cubic ($Pm\bar{3}m$) structure to a tetragonal structure (*P4/mmm*) with no octahedral tilts (*Figure 5*), denoted as a$^0$a$^0$c$^0$ [45] in Glazer notation [46].

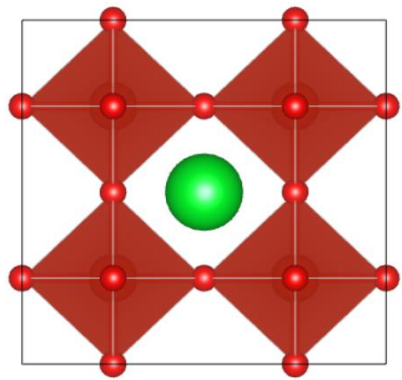

*Figure 5: The strained tetragonal (P4/mmm) SVO with no octahedral tilts*

The strain induced tetragonal reshaping of the unit cell lowers the point group symmetry from $O_h$ to $D_{4h}$ which causes a strain induced splitting of the bands at Γ and R points. As a result of this splitting, the threefold degeneracy previously observed in unstrained SVO at Γ and R points are thus lifted (*Figure 6.a-b*) [17,18]. This in turn results into a narrower band separation in tetragonal SVO irrespective of the type of strain applied.

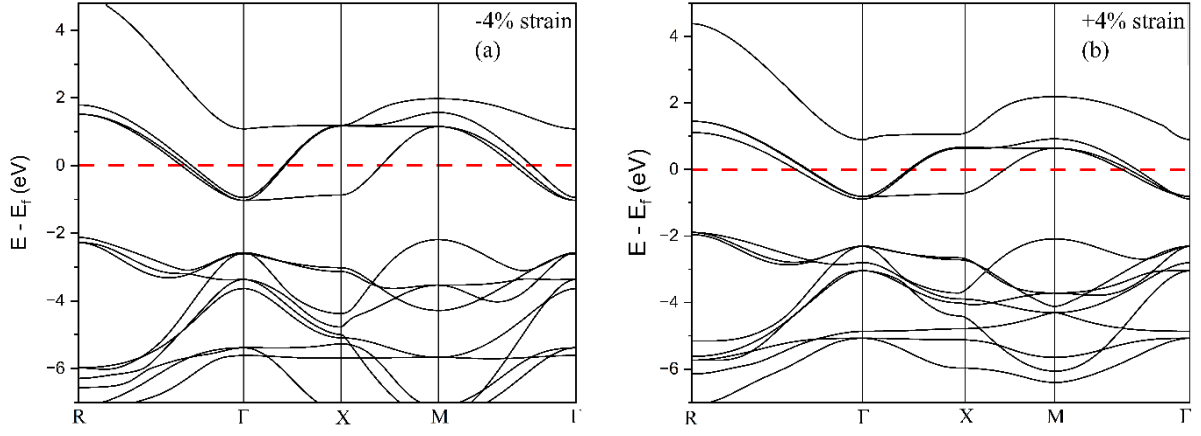

*Figure 6: Lifting of threefold degeneracy at band edges under 4% (a) compressive (b) tensile strain*

From Figure *6.(a-b)*, it is evident that at Γ(CBM) the $t_{2g}$ band is split into a single band ($d_{xy}$) and a pair ($d_{yz}$, $d_{zx}$) under both kinds of strain. Splitting of the $t_{2g}$ band is believed to favor a relatively insulating phase by orbital polarization [7,17,18] under both compressive and tensile strain. In case of tensile strain, the narrowing of the *d*-band (*Section 3.2*) reinforces this less conducting phase which causes the decrease in carrier concentration and increase in resistivity. In case of compressive strain, the *d*-band widening predominates, enhancing the metallic behavior with increased carrier concentration and decreased resistivity as observed in our calculations (*Section 3.3* and *3.4*). These two factors collectively explain the trend observed in plasma frequency as well (*Section 3.5*).

## 3.7 Variations in band separation ($E_{bs}$):

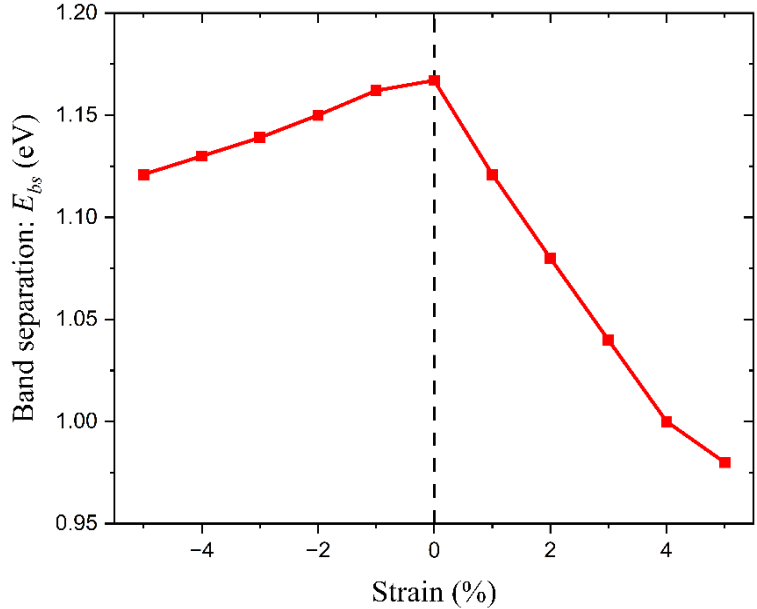

*Figure 7: Strain induced alterations in band separation ($E_{bs}$) due to the combined effects*

The band separation values ($E_{bs}$) between the conductions and valence bands shown in *Figure 7* are the results of the combined effects of band splitting and variation in bonding energy as explained below.

According to the linear combination of atomic orbital theory, the energy gap between the bonding and antibonding states is related to the separation ($E_{bs}$) between the conductions and valence bands. $E_{bs} = 2A + 2B$, $A$ and $B$ denote the atomic energy difference and bonding energy respectively [44]. This bonding energy can be defined as the stabilization and de-stabilization energies of the bonding and antibonding states respectively with respect to the corresponding atomic orbitals. For all the strained structures, the value of $A$ remains constant as the electronegativity difference between V and O is independent of the strain value, and $B$ is directly proportional to $V_{nn}$ ($V_{nn}$ – strength of nearest neighbor coupling) and hence to $1/d^2$ as $V_{nn} \propto 1/d^2$ ($d$ - nearest neighbor distance). Here $d$ essentially denotes the nearest distance between V and the adjacent O atom (*Table 1*).

Under tensile strain $d$ increases, thereby decreasing the atomic orbital overlap and nearest neighbor coupling and thus the bonding energy ($B$) decreases. As a result, the energy difference between the bonding and anti-bonding states decreases causing narrowing of band separation as observed in *Figure 7*. An exactly opposite trend is observed in case of compressive strain where band separation widening takes place. For biaxial strain, though the out-of-plane V-O bond length ($d$) is increased due to tensile strain, but the decrease in the V-O bond length along the other two in-plane directions predominate and cause band separation narrowing.

For compressive strains, the increase in the nearest neighbor coupling energy $V_{nn}$ widens the band separation whereas the degeneracy lifting narrows the band separation, causing a competing effect between the two. However, the band splitting effect predominates and thus band separation narrowing is observed in case of compressive strain as well, although less pronounced than tensile strain. Similar trends in strain induced band separation narrowing due to symmetry distortion had been observed previously by *Jeffrey B. Neaton et. al* in $SrTiO_3$ as well [10].

## 4. Conclusion:

First principles calculations based on density functional theory + Hubbard U are carried out to investigate the effect of strain on the electronic, transport and optical properties of the correlated oxide $SrVO_3$. Our calculations reveal that although strain does not alter the properties of SVO greatly, strain induced variations in various properties of SVO is indeed possible, especially to tune its transparency window. Biaxial strain is observed to lower the point group symmetry from $O_h$ to $D_{4h}$ and causes band splitting at the CBM(Γ) and VBM(R) which in turn reduces the band separation ($E_{bs}$). Under tensile strain, the $d$-band narrowing enhances electron-electron correlations and results in a decrease in carrier concentration and increase in resistivity. While in case of compressive strain, among $d$-band widening and degeneracy lifting which have competing effect, $d$-band widening dominates, resulting in an increase in carrier concentration and decrease in resistivity. Effect of strain in tuning the plasma frequency was also studied and its dependence on electron correlations and carrier concentration was investigated. It was observed that, under compressive strain, the in-plane plasma frequency value increases, limiting the transparency window while it decreases under tensile strain. It was apparent that for the applications of SVO as a transparent conductor, and to modulate the transparency window, strain tuning can be an effective way to obtain the desired efficiency. In this regard, balancing the trade-off between the competing phenomena present in the correlated oxide is necessary. Our calculations also provide insight on how band narrowing and relative position of the *d*-band center are related to the changes observed in transport and optical properties of SVO under strain. The results obtained here and the interpretations provided may further be useful for other materials sharing similar properties as SVO.